\documentclass[doublespacing]{elsart2}
\usepackage{bbm}
\usepackage{amsmath,amssymb,amsfonts,epsf}
\usepackage{graphicx}

\usepackage{lineno}

\usepackage{amssymb}

\begin{document}

\begin{frontmatter}

\title{Negro and Danube are mirror rivers}

\author[FEUP]{\corauthref{cor1}R. Gon\c calves}
\ead{rjasg@fe.up.pt}
\author[DMUM]{A. A. Pinto}

\corauth[cor1]{Universidade do Porto, R. Dr. Roberto Frias,
4200-465, Porto Portugal, tel/Fax: +351 225081707/1921}

\address[FEUP]{Faculdade de Engenharia da Universidade do Porto\\
R. Dr. Roberto Frias, 4200 - 465, Porto, Portugal\\}

\address[DMUM]{Universidade do Minho\\
4710 - 057, Campus de Gualtar, Braga, Portugal\\}

\begin{abstract}

We study the European river Danube and the South American river
Negro daily water levels. We present a fit for the Negro daily water
level period and standard deviation. Unexpectedly, we discover that
the river Negro and Danube are mirror rivers in the sense that the
daily water levels fluctuations histograms are close to the BHP and
reversed BHP, respectively.
\end{abstract}

\end{frontmatter}

\section{Introduction}
\label{sec:Int}

J\'anosi and Gallas \cite{ImreJason99} analyzed statistics of the
Alpine river Danube daily water level collected, over the period
1901-97, at Nagymaros, Hungary. The authors found, in the one day
logarithmic rate of change of the river water level, similar
characteristics to those of company growth (see Stanley et al.
\cite{Stanleyetal}) which shows that the properties seen in company
data are present in a wider class of complex systems. Bramwell et
al. \cite{bramwellfennelleuphys2002} defined a daily water level
mean and variance and computed the daily river water level
fluctuations. They have shown a data collapse of the Danube daily
water level fluctuations histogram to the (reversed)
Bramwell-Holdsworth-Pinton (BHP) probability density function (pdf).
Dahlstedt and Jensen \cite{DahlstedtJensen2005} described the
statistical properties of several river systems. They did a careful
study of the size of basin areas influence in the data collapse of
the rivers water level and runoff, in particular of river Negro at
Manaus, to the reversed BHP and to the Gaussian pdf showing that not
all rivers have the same statistical behavior. In this paper, we
study, again, the South American river Negro daily water level at
Manaus (104 years). We compute and present a cyclic fit for the
Negro daily water level period and the Negro daily water level
standard deviation. We show that the histogram of the Danube water
level fluctuations is on top of the reversed BHP pdf, which does not
happen for the Negro daily water level.

\section{BHP and the Danube and Negro data}
\label{sec:BHPDanubeDurius}


We define the \emph{Danube daily water level period}
$\hat{l}_\mu(t)$ by

\begin{equation}
         \hat{l}_\mu(t)=\frac{1}{T}\sum_{j=0}^{T-1} X(t+365j) ,
        \label{eq1}
\end{equation}

\noindent where $T=87$ is the number of observed years and $X_t$ is
the Danube daily water level time series.\footnote{All the
observations of the days $29^{\mbox{th}}$ of February were
eliminated.}

\begin{figure}[!htb]
\begin{center}
\includegraphics[width=12cm]{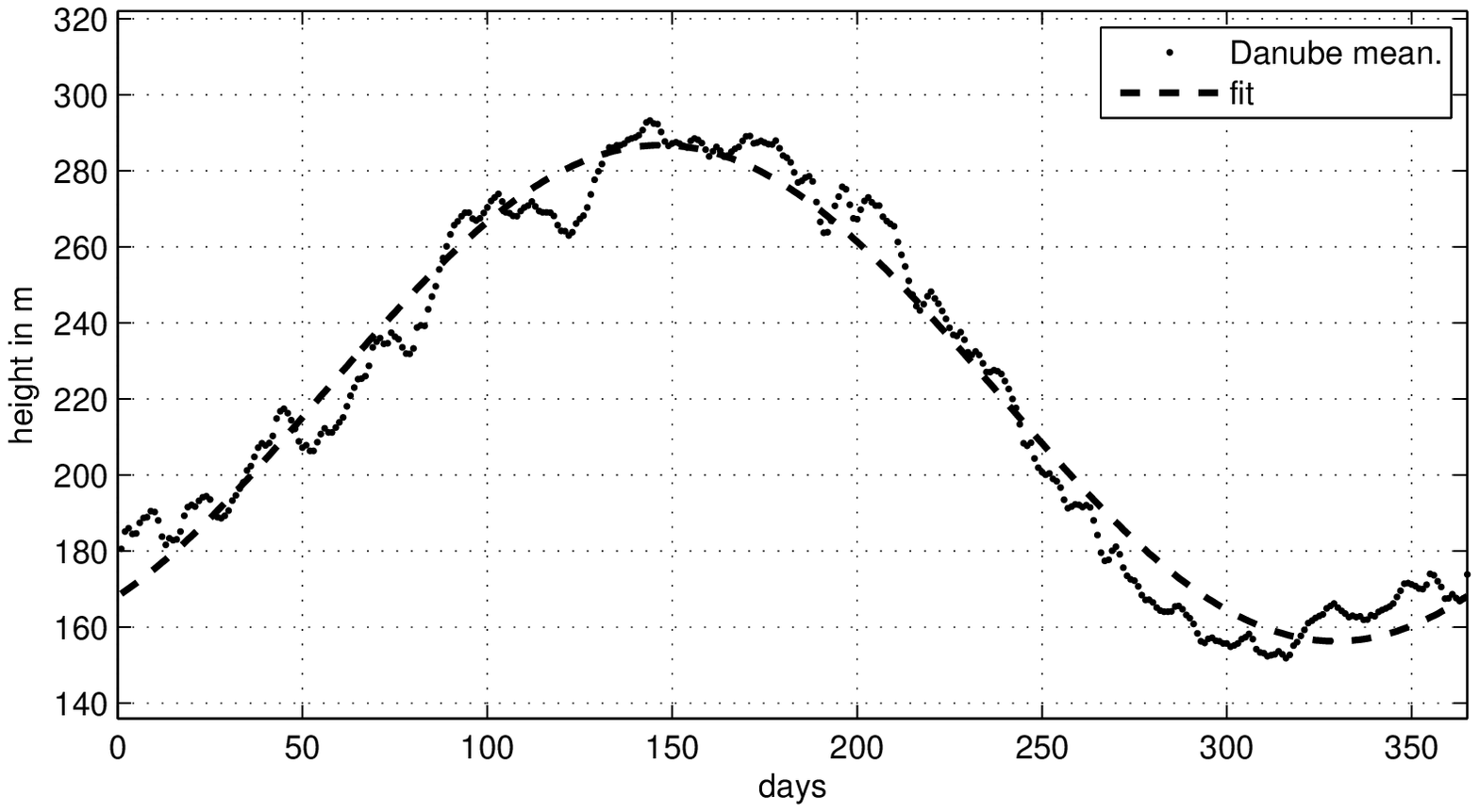}
\caption{\footnotesize{Chronograph of the Danube daily water level
period $\hat{l}_{\mu}(t)$, in a semi-log plot.}} \label{fig1}
\end{center}
\end{figure}

In Figure \ref{fig1}, we show a fit to the Danube daily water level
period $\hat{l}_\mu(t)$. The mean period fit, using the first
harmonic of the Fourier series, is given by

\begin{equation}
 \tilde{l}_\mu(t)=443.06-53.3409\sin\left(\frac{2t\pi}{365}\right)+37.4623\cos\left(\frac{2t\pi}{365}\right)
 \ .
\label{eq2}
\end{equation}
The percentage of variance explained by the fit is $R^2=99.8\%$.


We define the \emph{Negro daily water level period} $\hat{n}_\mu(t)$
by

\begin{equation}
         \hat{n}_\mu(t)=\frac{1}{T}\sum_{j=0}^{T-1} Y(t+365j) ,
        \label{eq3}
\end{equation}
\noindent
 where $T=104$ is the number of observed years and $Y_t$ is the Negro daily water level time series.
\footnote{All the observations of the days $29^{\mbox{th}}$ of
February were eliminated.}

\begin{figure}[!htb]
\begin{center}
\includegraphics[width=12cm]{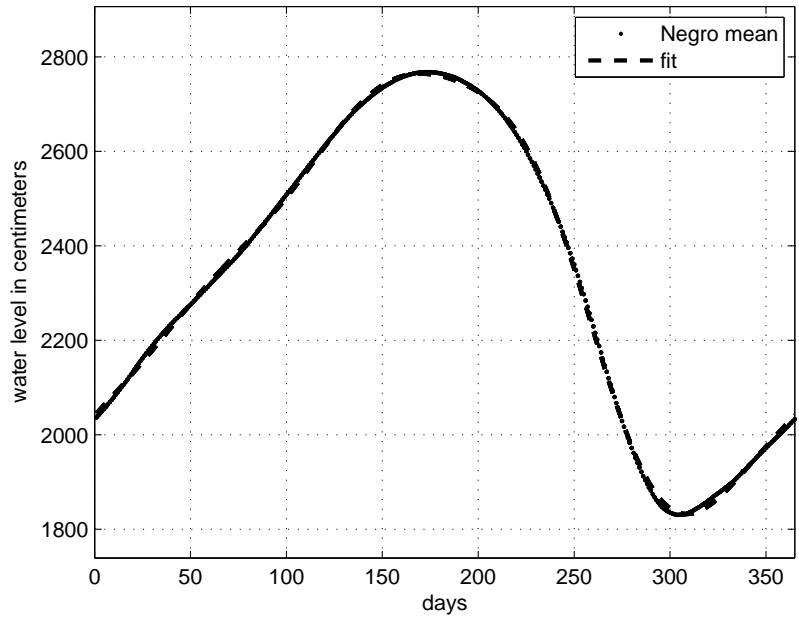}
\caption{Chronogram of the Negro daily water level period
$\hat{n}_\mu(t)$ and fit $\tilde{n}_\mu(t)$.} \label{fig2}
\end{center}
\end{figure}

In Figure \ref{fig2}, we show a fit $\tilde{n}_\mu(t)$ of the Negro
daily water level period $\hat{n}_\mu(t)$. The mean period fit,
using the first four sub-harmonics of the Fourier series, is given
by

\begin{equation}
 \tilde{n}_\mu(t)=\frac{a_0}{2}+\sum_{n=1}^{4}{a_n\cos\left(\frac{2tn\pi}{365}\right)+b_n\sin\left(\frac{2tn\pi}{365}\right)}\quad.
\label{eq4}
\end{equation}
where $a_n$ and $b_n$ are given in table \ref{tab1}. The percentage
of variance explained by the fit is  $R^2=99.9\%$.

We define the \emph{Danube daily water level standard deviation}
$\hat{l}_\mu(t)$ by

\begin{equation}
         \hat{l}_{\sigma}(t)=\sqrt{\frac{\sum_{j=0}^{T-1} X(t+365j)^2}{T}-{\hat{l}_\mu(t)}^2}\quad.
        \label{eq5}
\end{equation}

\begin{figure}[htb]
\begin{center}
\includegraphics[width=12cm]{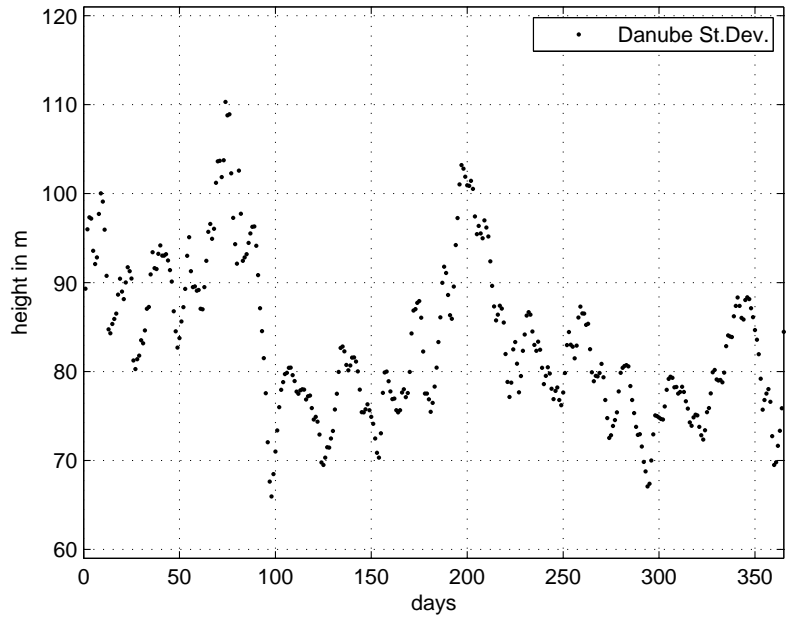}
\caption{\footnotesize{Chronogram of the Danube daily water level
standard deviation $\hat{l}_\sigma(t)$}} \label{fig3}
\end{center}
\end{figure}

In Figure \ref{fig3}, we show the chronogram of the Danube daily
water level standard deviation.

We define the \emph{Negro daily water level standard deviation}
$\hat{n}_{\sigma}(t)$ given by

\begin{equation}
         \hat{n}_{\sigma}(t)=\sqrt{\frac{\sum_{j=0}^{T-1} Y(t+365j)^2}{T}-{\hat{n}_\mu(t)}^2}\quad.
        \label{eq6}
\end{equation}

\begin{figure}[!htb]
\begin{center}
\includegraphics[width=12cm]{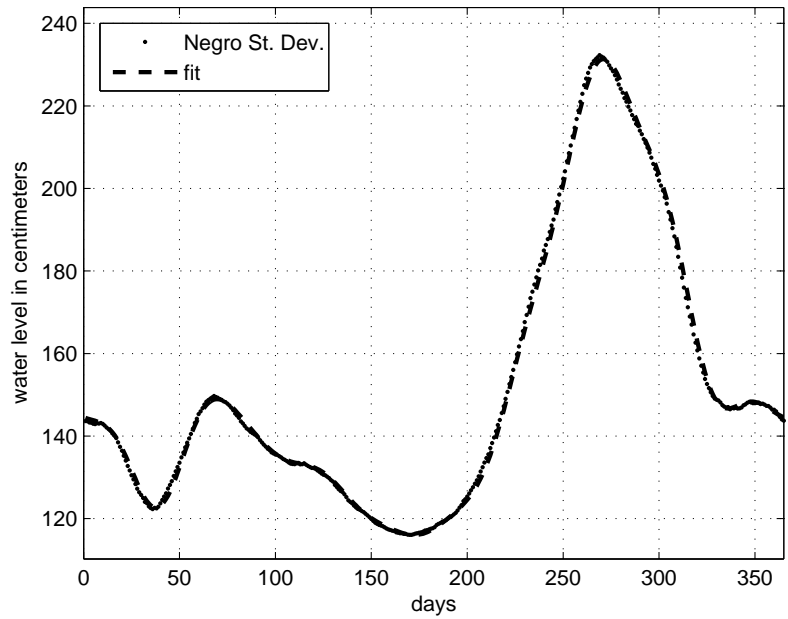}
\caption{\footnotesize{Chronogram of the Negro daily water level
standard deviation $\hat{n}_\sigma(t)$} and fit
$\tilde{n}_\sigma(t)$.} \label{fig4}
\end{center}
\end{figure}

In Figure \ref{fig4}, we show a fit of the Negro daily water level
standard deviation. The fit $\hat{n}_\sigma$, using the first ten
sub-harmonics of the Fourier series, is given by

\begin{equation}
 \tilde{n}_\sigma(t)=\frac{a_0}{2}+\sum_{n=1}^{10}{a_n\cos\left(\frac{2tn\pi}{365}\right)+b_n\sin\left(\frac{2tn\pi}{365}\right)}
\label{eq7}
\end{equation}
where $a_n$ and $b_n$ are given in table \ref{tab2}. The percentage
of variance explained by the fit  is  $R^2=88.6\%$.



Following Bramwell et. al \cite{bramwellfennelleuphys2002}, we
define the \emph{Danube daily water level fluctuations} $l_f(t)$ by

\begin{equation}
          l_f(t)=\frac{l(t)-\hat{l}_\mu(t)}{\hat{l}_\sigma(t)}\quad.
        \label{rjasgeq3}
\end{equation}

\begin{figure}[!htb]
\begin{center}
\includegraphics[width=12cm]{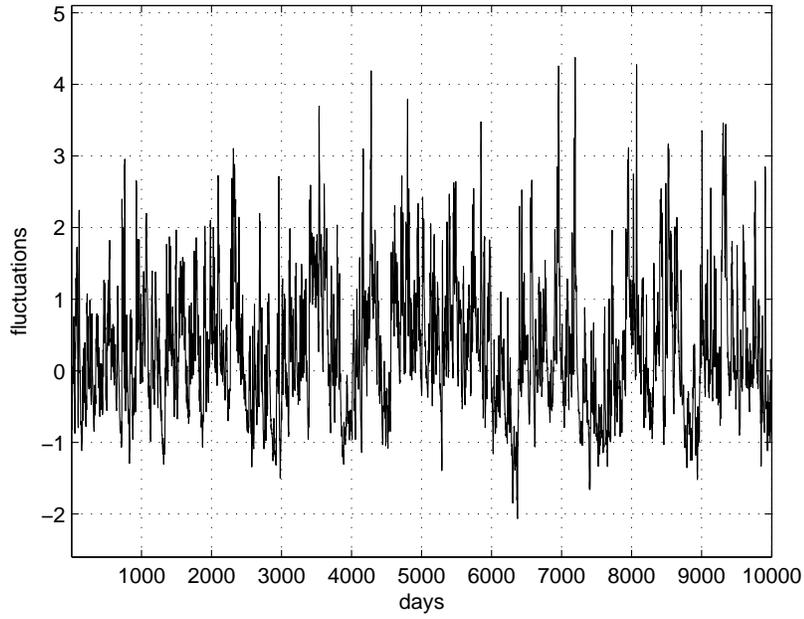}
\caption{Chronogram of the Danube daily water level fluctuations
$l_f(t)$.} \label{fig5}
\end{center}
\end{figure}

\begin{figure}[!htb]
\begin{center}
\includegraphics[width=12cm]{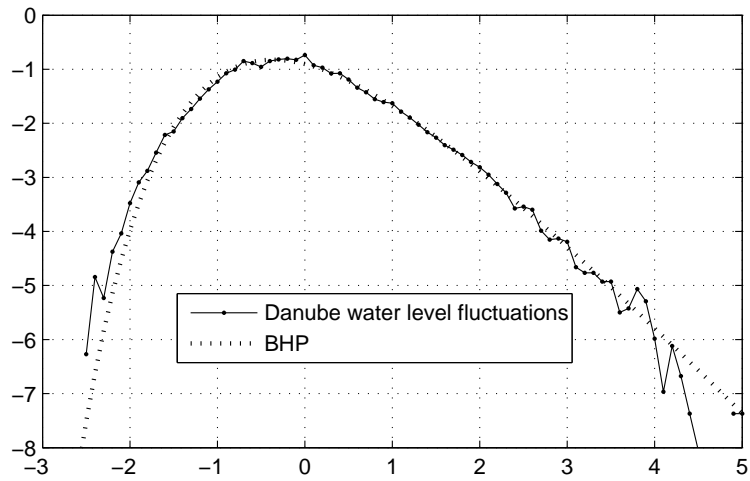}
\caption{\footnotesize{Data collapse of the histogram of the Danube
water level fluctuations to the BHP pdf, in the semi-log scale.}}
\label{fig7}
\end{center}
\end{figure}

In Figure \ref{fig5}, we show the Danube daily water level
fluctuations $l_f(t)$. As shown by Bramwell et al.
\cite{bramwellfennelleuphys2002}, the (reversed) BHP pdf falls on
top of the histogram of the Danube daily water level fluctuations,
in the semi-log scale (see Figure \ref{fig7}).


We define the \emph{Negro daily water level fluctuations} $n_f(t)$
by

\begin{equation}
          n_f(t)=\frac{n(t)-\hat{n}_\mu(t)}{\hat{n}_\sigma(t)}\quad.
        \label{rjasgeq3}
\end{equation}
\noindent

\begin{figure}[!htb]
\begin{center}
\includegraphics[width=12cm]{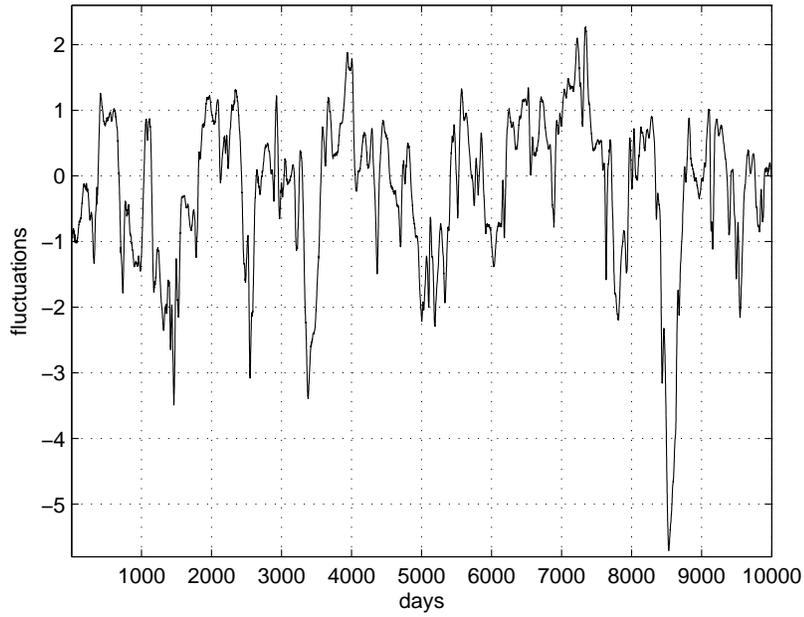}
\caption{Chronogram of the Negro daily water level fluctuations
$n_f(t)$.} \label{fig6}
\end{center}
\end{figure}

\begin{figure}[!htb]
\begin{center}
\includegraphics[width=12cm]{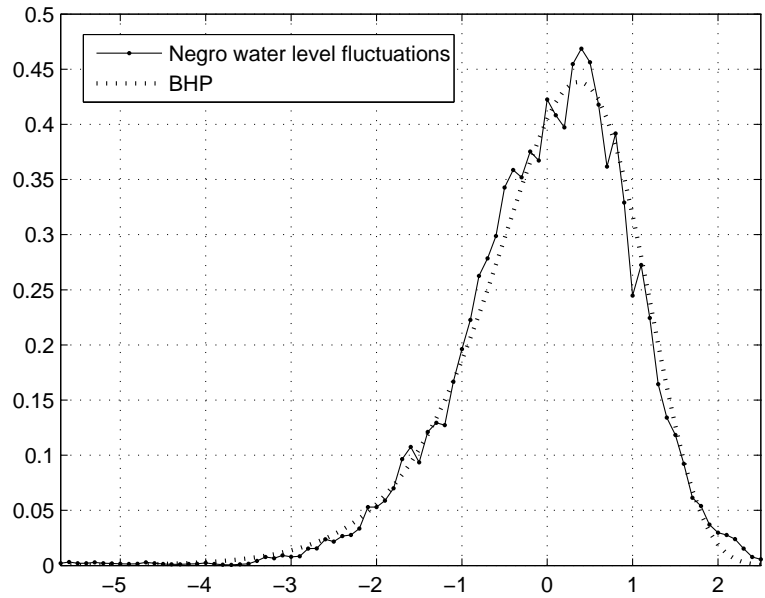}
\caption{Histogram of the Negro daily water level fluctuations with
the BHP on top.} \label{fig8}
\end{center}
\end{figure}

\begin{figure}[!htb]
\begin{center}
\includegraphics[width=12cm]{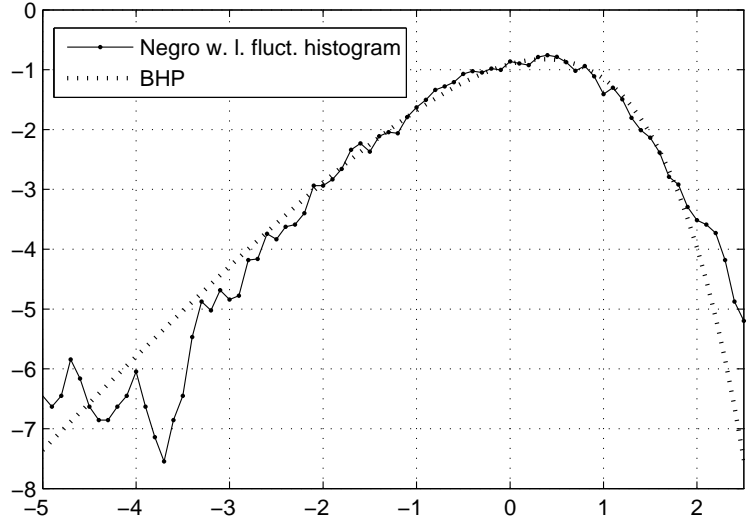}
\caption{Histogram of the Negro daily water level fluctuations for
the full year, in the semi-log scale, with the BHP on top.}
\label{fig9}
\end{center}
\end{figure}

 In Figure \ref{fig6}, we show the Negro daily water level
fluctuations $n_f(t)$. In Figure \ref{fig8}, we show the histogram
of the Negro daily water level fluctuations. In Figure \ref{fig9},
we show the data collapse of the histogram in the semi-log scale to
the BHP pdf.

\clearpage

\section{Conclusions}

We computed and presented a cyclic fit for the Negro daily water
level period and for the Negro daily water level standard deviation
using the first four and ten sub-harmonics, respectively.
 We computed the histogram of the Negro daily water level
fluctuations at Manaus, and we compared it with the BHP pdf. The
histogram of the Negro daily water level fluctuations is close to
the BHP pdf. We have shown that the histogram of the Danube water
level fluctuations is on top of the reversed BHP pdf, which does not
happen for the river Negro daily water level fluctuations.


\section*{Acknowledgements}
We thank Imre J\'anosi for providing the river Danube data and Mrs.
Andrelina Santos of the Ag\^encia Nacional de \'Aguas of Brazil for
providing the river Negro data.

\clearpage
\noindent

\noindent

\noindent

\noindent

\noindent

\noindent

\noindent

\noindent

\noindent

\begin{table}[!htb]
 \centering
  \caption{\footnotesize{Fourier Coefficients for the Negro
daily water level period $\hat{n}_\mu(t)$.}} \label{tab1}
\newcommand{\m}{\hphantom{$-$}}
\newcommand{\cc}[1]{\multicolumn{1}{c}{#1}}
\renewcommand{\tabcolsep}{1pc} 
\renewcommand{\arraystretch}{1.0} 
\begin{tabular}{@{}lll}\hline
 n &  $a_n$    &   $b_n$\\
 0  &  4671.2    &    -\\
 1  & -386.92  &     195.28\\
 2  &  73.672  &     74.577\\
 3  &  29.0336  &    -12.473\\
 4  & -9.9726  &    -13.724\\
\hline
\end{tabular}
\end{table}
\noindent

\begin{table}[htb]
\caption{\footnotesize{Fourier Coefficients for the Negro daily
water level standard deviation $\hat{n}_\sigma(t)$.}}
 \label{tab2}
 \centering
\newcommand{\m}{\hphantom{$-$}}
\newcommand{\cc}[1]{\multicolumn{1}{c}{#1}}
\renewcommand{\tabcolsep}{1pc} 
\renewcommand{\arraystretch}{1.0} 
\begin{tabular}{@{}lll}
\hline
n   &  $a_n$    &   $b_n$\\
0   &  304.1   &    -\\
1   &   -8.505  &   33.983  \\
2   &   -28.086 &   1.1481  \\
3   &   -2.6941 &   -8.1938 \\
4   &   6.5597  &   -1.9674 \\
5   &   -3.6529 &   1.6374  \\
6   &   2.1779  &   1.2313  \\
7   &   -0.2847 &   -1.2559 \\
8   &   -0.5713 &   -0.6129 \\
9   &   1.7276  &   0.54441 \\
10  &   -1.0358 &   0.64293 \\
\hline
\end{tabular}
\end{table}

\end{document}